\documentstyle[aps,twocolumn,epsfig]{revtex}

\begin{document}
\draft
\title{Direct photons from relativistic heavy ion collisions at CERN SPS and
at RHIC}
\author{{\bf A. K. Chaudhuri\cite{byline}}}
\address{Department of Physics, Technion-Israel Institute of Technology,\\
Haifa 32000, Israel\\
and}
\address{ Variable Energy Cyclotron Centre\\
1/AF,Bidhan Nagar, Calcutta - 700 064\\
}
\author{{\bf T. Kodama}}
\address{Instituto de F\'{\i}sica, Universidade Federal do Rio de Janeiro\\
Cxp 68528, Rio de Janeiro, 21945-970, RJ, Brazil}
\maketitle

\begin{abstract}
Assuming QGP as the initial state, we have analyzed the direct photon data,
obtained by the WA98 collaboration, in 158 A GeV Pb+Pb collisions at CERN
SPS. It was shown, that for small thermalisation time, two loop rate
contribute substantially to high $p_T$ photons. We argue that for extremely
short thermalisation time scale, the higher loop contribution should not be
neglected. For thermalisation time 0.4 fm or greater, when higher loop
contribution are not substantial, the initial temperature of the QGP is not
large and the system does not produce enough hard $p_T$ photons to fit the
WA98 experiment. For initial time in the ranges of 0.4-1.0 fm, WA98 data
could be fitted only if the fluid has initial radial velocity in the range
of 0.3-0.5c. The model was applied to predict photon spectrum at RHIC energy.
\end{abstract}




Ever since possible existence of deconfining phase transition from hadrons
to quark gluon plasma (QGP) was predicted by the Quantum Chromodynamics
(QCD), nuclear physicists are eagerly looking for the experimental evidences
of QGP, the new state of matter\cite{QM2001}. It has even been claimed that,
evidence of the new state of matter is already seen at CERN SPS in Pb+Pb
collision \cite{he00}. The claim was based mainly on the experimental $%
J/\psi $ suppression and strangeness enhancement, in Pb+Pb collisions. Due
to intrinsic ambiguities in disentangling the similar effects arising from
hadronic processes, their claim is rather considered controversial\cite{Zsch}%
. On the other hand, it may be noticed that electromagnetic signals
(dileptons and photons), were not considered in establishing the claim, even
though, photons and dileptons are well recognized probes of QGP. They suffer
minimal final state interactions and can give information about the early
stage of the collisions. Experimentally however, it is a challenge to obtain
precision photon data from huge background. Recently WA98 collaboration has
published their single photon emission data for 158 A GeV Pb+Pb collisions
at CERN SPS \cite{wa98}. Several authors have analyzed the WA98 single
photon data \cite{sr00,ja00,ga00,pe00,ch01}, but their conclusions are not
convergent. It seems that the data could be explained in a QGP or in a
hadronic gas scenario, with or without initial radial velocity.

In the present paper, we further analyze the WA98 data in the initial QGP
scenario. We first assume that the thermalisation time $\tau _{i} $, beyond
which hydrodynamics is applicable, is larger than 0.4 fm. The reason for
this is the uncertainty in the elementary photon production rate. For
smaller $\tau _{i}$, higher loops contributions can be substantial, ruining
any agreement obtained with data. If $\tau _{i}>0.4$ fm is the case, we then
confirm the analysis of Peressounko et al \cite{pe00} that the data require
an initial radial velocity. It was also shown that the high $p_{T}$ photon
spectra depend quite strongly on the initial radial velocity distribution.
For realistic (surface peaked) velocity distribution, initial radial
velocity in the range of 0.3-0.4c is required.

Procedure for obtaining photon spectra in a hydrodynamic evolution is well
known \cite{ge86}. We have solved the hydrodynamic equations $\partial _{\mu
}T^{\mu \nu }=0$ for a baryon free gas assuming cylindrical symmetry and
boost-invariance. The equation of state for QGP was assumed to be $%
p_{q}=a_{q}T^{4}-B$ with $a_{q}=42.25\pi ^{2}/90$. The hadronic equation of
state was generalized to include all the mesonic resonances with mass $<$ 2
GeV. The cut off 2 GeV is rather arbitrary and we verify that the results do
not depend on the value of cut-off significantly. The bag constant $B$ was
obtained from the Gibbs condition $p_{QGP}(T_{c})=p_{had}(T_{c})$. The
critical temperature ($T_{c}$) and the freeze-out temperature ($T_{F}$) were
assumed to be 180 MeV and 100 MeV respectively.

Hydrodynamic models require, as inputs, the initial time ($\tau _{i}$) and
initial energy density or equivalently the temperature ($T_{i}$) of the
fluid. $\tau _{i}$ is the thermalisation time, beyond which
quasi-equilibrium is established and hydrodynamics is applicable. $\tau _{i}$
and $T_{i}$ are parameters of the model, unless one obtains them from some
microscopic transport models. In order to reduce parameters, one further
assume that the fluid flow is isentropic. Then for a given $\tau _{i}$ the
initial temperature $T_{i}$ of the fluid (QGP) can be obtained by relating
the entropy density with the observed pion multiplicity (assuming pion
decoupling to be adiabatic) \cite{hw85},

\begin{equation}  \label{1}
T^3_i\tau_i=\frac{1}{\pi R^2_A} \frac{c}{4a_{q,h}}\frac{dN}{dY} (b=0)
\end{equation}

\noindent where $c=2\pi^4/45\zeta(3)$ and $R_A$ is the transverse radius of
the system (assumed to be 6.4 fm for Pb+Pb collisions). $b=0$ corresponds to
central collisions. In table 1, initial temperature of the QGP, for certain
values of the initial time $\tau_i$ are shown. Rapidity density was assumed
to be $dN/dY$=750 \cite{sr00}. It can be seen that choice of initial time is
of crucial importance. QGP is formed at higher temperature for low values of
the thermalisation time. High $p_T$ photon production rate depends strongly
on the initial temperature and it is possible to increase its yield by
arbitrarily reducing the thermalisation time and correspondingly increasing
the initial temperature of the QGP fluid.

For the single photons from hadronic gas we include the following processes,

(a) $\pi\pi \rightarrow \rho \gamma$, (b) $\pi \rho \rightarrow \pi \gamma$,
(c) $\omega \rightarrow \pi \gamma$, (d) $\rho \rightarrow \pi \pi \gamma$
(e) $\pi \rho \rightarrow A_1 \rightarrow \pi \gamma$

\noindent rates for which are well known \cite{na92,xi92}.

Rate of production of hard photons from QGP were evaluated by Kapusta et al 
\cite{ka91}. To one loop order,

\begin{equation}
E\frac{dR}{d^{3}p}=\frac{1}{2\pi ^{2}}\alpha \alpha
_{s}\sum_{f}e_{f}^{2}T^{2}e^{-E/T}\ln (\frac{\xi E}{\alpha _{s}T})  \label{2}
\end{equation}

\noindent where $\xi \sim 0.23$ is a constant. The summation runs over the
flavors of the quarks and $e_{f}$ is the electric charge of the quarks in
units of charge of the electron.

Recently Aurenche et al\cite{au98} evaluated the production of photons in a
QGP. At two loops level Bremsstrahlung photons $(qq(g) \rightarrow qq(g)
\gamma$ found to be dominating the compton and annihilation photons.


The rate of production of photons due to Bremsstrahlung (corrected for the
factor of 4 \cite{au98a,st01}) was evaluated by them as,

\begin{equation}  \label{3}
E \frac{dR}{d^3p} =\frac{1}{4} \frac{8}{\pi^5} \alpha \alpha_s \sum_f e^2_f 
\frac{T^4}{E^2} e^{-E/T} (J_T - J_L) I(E,T)
\end{equation}

\noindent where $J_T \sim 4.45$ and $J_L \sim -4.26$ for two flavors and 3
colors of quarks. For 3 flavor quarks, $J_T \sim 4.8$ and $J_L \sim -4.52$. $%
I(E,T)$ stands for,

\begin{eqnarray}  \label{4}
I(E,T) = &&[3\zeta(3) + \frac{\pi^2}{6}\frac{E}{T} +(\frac{E}{T})^2 \ln 2 +4
Li_3(-e^{-|E|/T})  \nonumber \\
&&+ 2Li_2(-e^{-|E|/T}) -(E/T)^2 \ln(1+e^{-|E|/T})]
\end{eqnarray}

\noindent and the poly-logarith functions $Li$ are given by,

\begin{equation}  \label{5}
Li_a(z) = \sum_{n=1}^{\infty} \frac{z^n}{n^a}
\end{equation}

Aurenche et al \cite{au98} also calculated the contribution of the $q\bar{q}$
with scattering, which should also be corrected for the same factor of 4.
The corrected rate is,

\begin{equation}  \label{6}
E \frac{dR}{d^3p} =\frac{1}{4} \frac{8}{3\pi^5} \alpha \alpha_s \sum_f e^2_f
E T e^{-E/T} (J_T - J_L)
\end{equation}

We would like to call attention to the fact that two-loop photon rate from
QGP is not complete. Higher loops contribute to the same order \cite{au00a}.
Also Landau-Migdal-Pomeranchuk effect has been neglected \cite{au00b}.
Questions naturally arise how can then one obtain a reliable estimate of
photon yield in the QGP scenario. In Fig.1, we have shown the ratio of
photon yield obtained with one+two loop rate and one loop rate, for initial
times $\tau _{i}$ =0.2,0.4,0.6,0.8 and 1 fm. It can be seen for the lowest
thermalisation time ($\tau _{i}$=0.2 fm) two loop rate contribute more that
70\% to the high $p_{T}$ photons. If the higher loops contribute to the same
order as the two-loop rate, their inclusion can change the two loop results
drastically. In the given circumstances, it will be unwise to apply these
results if the thermalisation time scale is as small as 0.2 fm. 

\begin{figure}[h]
\centerline{\psfig{figure=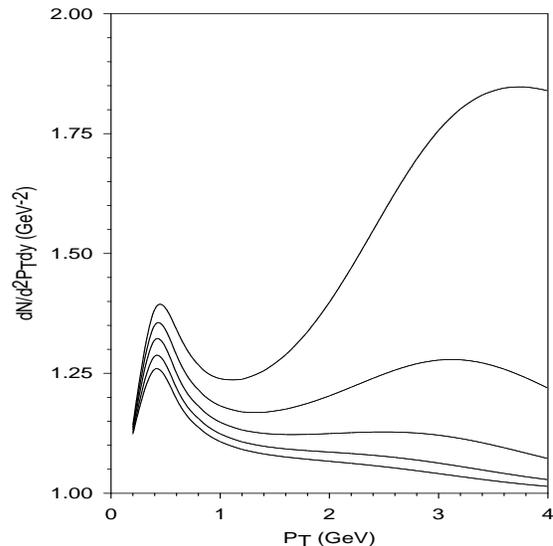,height=7.5cm,width=7.5cm}}
\caption{Ratio of photon yield in QGP scenario obtained with one+two loop
rate and one loop rate, for different initial times. The lines, from top to
bottom corresponds to $\protect\tau_i$=0.2, 0.4, 0.6, 0.8 and 1 fm.
Corresponding initial temperatures are listed in table 1. The lines are
drawn with }
\end{figure}

For $\tau _{i}$ $\geq $ .4 fm or more, the two loop rates contribute to 25\%
or less to the total yield. Thus for a realistic estimate of photon yield in
the QGP scenario, one must consider $\tau _{i}\geq $ .4 fm, when
contribution of two loop rate is moderate and even if higher loops
contribute to the same order, it will not affect the total yield
substantially. Calculations with smaller $\tau _{i}$ could not be relied
upon. We thus assume $\tau _{i}\geq $ 0.4fm. As will be shown later, for $%
\tau _{i}\geq $ 0.4fm, initially static QGP scenario could not explain the
WA98 data. It under predicts them. However, if the fluid is assumed to have
certain initial radial velocity, data could be fitted satisfactorily.
Indeed, Peressounko et al \cite{pe00} argued that the WA98 data could be
fitted only with some initial radial velocity. Present analysis confirms the
result. Source of initial radial velocity could be the collisions among the
constituents of the fluid (quarks and gluons). However, the high $p_{T}$
photon yield will depend on the type of velocity distribution to be assumed.
Peressounko et al \cite{pe00} used linearly increasing velocity
distribution. In an earlier work \cite{ch01} a Woods-Saxon form for the
initial velocity distribution was used. Fig.2, we have shown the computed
photon yield for three types of velocity distributions,

\noindent (a) Woods-Saxon type of velocity: 
\begin{equation}
v_{r}(r)=v_{r}^{0}/(1+exp((r-R_{0})/a)
\end{equation}

\noindent (b) Linearly increasing velocity:

\begin{equation}
v_{r}(r)=\left\{ 
\begin{array}{c}
v_{r}^{0}r/R_{0}, \\ 
0%
\end{array}%
\begin{array}{c}
r\leq R_{0} \\ 
r>R_{0}%
\end{array}%
\right.
\end{equation}

\noindent and (c) surface peaked velocity:

\begin{equation}
v_{r}(r)=4v_{r}^{0}exp((r-R_{0})/a)/(1+exp((r-R_{0})/a))^{2}
\end{equation}

\noindent with $R_{0}=$6.4 fm and $a=$0.54 fm corresponding to Pb nucleus.
In these velocity profiles, the parameter $v_{r}^{0}$ is chosen to be the
maximum value of the distribution. We should remember that this choice of
parametrization is not unique. For example, it may also be normalized to
give the same average radial velocity. Due to the ambiguity of the
definition of $v_{r}^{0}$, the absolute values of photon yield for each
profile in Fig. 2 may change in different normalization, but an important
fact is that $p_{T}$ dependence changes and the high $p_{T}$ part of the
photon spectra is sensitive 
\begin{figure}[h]
\centerline{\psfig{figure=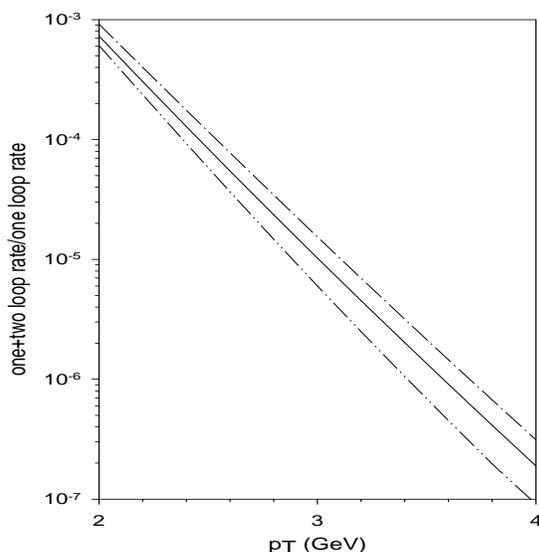,height=7.5cm,width=7.5cm}}
\caption{Photon yield for different types of initial velocity distributions,
with $v_{r}^{0}$=0.2c. The dash-dot-dot line is for the surface peaked
initial velocity, the dash-dot line is for the Woods-Saxon type of velocity
distribution and the solid line is for the linearly increasing velocity
profile. }
\end{figure}

\noindent to the shape of the velocity distribution. Surface peaked velocity
distribution is least effective for producing $p_{T}$. On the other hand,
Saxon-Woods type of velocity profile produces more high $p_{T}$ photons.
Linearly increasing velocity lies in between.

To see the physical picture of these profiles, let us consider some
conserving quantity associated to the fluid. Then from the continuity
equation for cylindrical symmetric case,%
\[
\dot{n}+\frac{1}{r}\frac{\partial }{\partial r}\left( rnv_{r}\right) =0,
\]%
where $n$ is the density of the conserved quantity. For example, if we take
Wood-Saxon type velocity field, at $r\simeq 0,$ $v_{r}\simeq const$, so that%
\[
\dot{n}\simeq -v_{r}\left( 0\right) \left[ \frac{n}{r}+\frac{\partial n}{%
\partial r}\right] 
\]%
near $r\simeq 0$. Except for $n\left( 0\right) \rightarrow 0$ for $%
r\rightarrow 0$, the density variation at the center becomes infinitely
large. Thus, the Wood-Saxon type profile represents a detonation at the
origin, exploding from the center whole the material outwords. The linear
profile corresponds to the homologous expansion of the system, corresponding
to a rather mild density change. For example, a typical solution is%
\[
n\simeq \frac{1}{R^{2}\left( t\right) }f\left( \frac{r}{R\left( r\right) }%
\right) ,
\]%
where $f$ is an arbitrary function. On the other hand, when the QGP gas is
initially almost at rest, and its surface area is expanding rapidly to the
vacuum, similar to the early simple wave stage of the Landau model, then the
velocity profile becomes surface peaked. If a nuclear collision generates
initially almost static QGP fireball, the fluid velocity may be more surface
peaked compared to the linear one.

Before we compare our model prediction with the WA98 experiment, we would
like to make few comments on the direct QCD and pre- equilibrium photons.
Direct QCD photons are produced from the early hard collisions of partons in
the nuclei and in Pb+Pb collisions make significant contribution to the high 
$p_{T}$ yield. Gallmeister et al \cite{ga00} claimed that prompt photons are
able to explain the high $p_{T}$ data in Pb+Pb collisions. Dumitru et al 
\cite{du01} also arrived at a similar conclusion including the nuclear
broadening effects. However, this point is still controversial due to
uncertainties in prompt photon emission at AA collisions. Thus Alam et al 
\cite{ja00} and also Srivastava and Sinha \cite{sr00} calculated the prompt
photon emission for Pb+Pb collisions. It was seen that for $p_{T}$ $>$2 GeV,
direct QCD photons alone can describe the data within a factor of 3-8 only.

Pre-equilibrium photons are emitted before the establishment of quasi
equilibrium. Traxler and Thoma \cite{tr96} calculated pre-equilibrium
photons and found them order of magnitude less than the equilibrium photons.
Roy et al \cite{ro97} also calculated the pre-equilibrium photons. They used
Fokker-Plank equations and found that pre-equilibrium photons are less or at
best equal to the equilibrium photons. The results refers to RHIC and LHC
energies. At SPS one expects still lower contribution of pre-equilibrium
photons. At SPS, the system is far from chemical equilibrium and number of
quarks and anti-quarks will be less. Pre-equilibrium photons then can be
neglected.

\begin{figure}[h]
\centerline{\psfig{figure=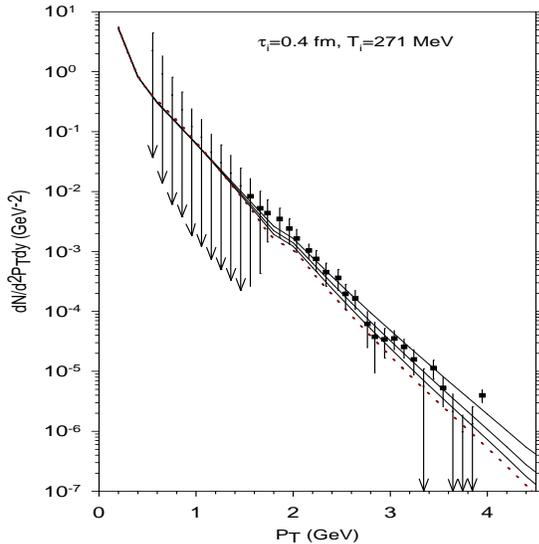,height=7.5cm,width=7.5cm}}
\caption{Thermal plus hard QCD photons compared with WA98 experimental
results. The initial time and temperature of QGP is 0.4 fm and 271 MeV
respectively. The solid lines are obtained with initial radial velocity
0.2,0.3 and 0.4c (from bottom to top). The dotted line is obtained without
any initial radial velocity. }
\end{figure}

In Fig.3 , we have compared the photon yield obtained in the present model
with WA98 experiment. The initial time and temperature are $\tau _{i}$=0.4
fm and $T_{i}$=271 MeV respectively. We have included the hard QCD photons
for $p_{T}$ $>$ 2 GeV following Alam et al \cite{ja00}. Without any initial
radial velocity (the dotted line) thermal photons together with hard QCD
photons could not explain the data. High $p_{T}$ part of the data are
underestimated by a factor of 5 or more. It is possible to obtain a
reasonable fit to data for still lower initial time(e.g. $\tau _{i}$=0.2 fm, 
$T_{i}$=341 MeV), but as emphasized earlier, the results obtained thus are
not reliable. Higher loops (three and more) contributions, which are
neglected here may then be significant. When we suppose that the initial
temperature is not so high, then the data requires an initial radial
velocity in our model. In Fig.3, the solid lines are the photon yield
obtained with initial radial velocity with maximum velocity $v_{r}^{0}$%
=.2,.3,.4c. Here we take the surface peaked initial velocity profile, which
corresponds to a scenario of QGP fireball with surface expansion. As
expected with initial $v_{r}$, high $p_{T}$ photon yield is increased and we
find that the data is well describe if $v_{r}^{0}$ lies in the range of
0.3-0.4c. When we assume more mild QGP gas expansion represented by the
linear velocity profile, then these values are little bit smaller. In Fig.4,
same results are shown for initial time $\tau _{i}$=1.0 fm. In this case,
without any initial radial velocity, data are more underpredicted.
Reasonable agreement with data is obtained again with (surface peaked)
initial velocity in the range of $v_{r}^{0}$=0.4-0.5c. The present results
are in agreement with the finding of Peressounko et al \cite{pe00} that the
WA98 single photon data require initial radial velocity of the fluid, at
least in the surface region.

\begin{figure}[h]
\centerline{\psfig{figure=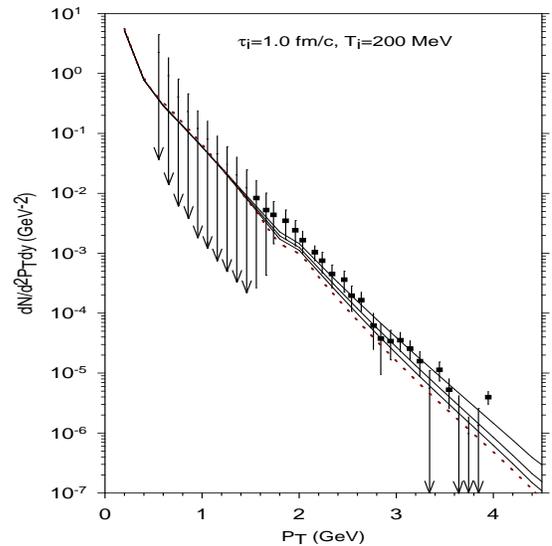,height=7.5cm,width=7.5cm}}
\caption{Thermal plus hard QCD photons compared with WA98 experimental
results. The initial time and temperature of QGP is 1.0 fm and 200 MeV
respectively. The solid lines are obtained with initial radial velocity 0.3,
0.4 and 0.5c (from bottom to top). The dotted line is obtained without any
initial radial velocity. }
\end{figure}

We now show the prediction for RHIC energy. In Fig.5, we have shown the
results obtained for Au+Au collisions at $\sqrt{s}$=130 GeV. The initial
time and temperature was assumed to be $\tau_i$=0.5 fm and $T_i$=300 MeV 
\cite{du99}. The dash-dot line is the hard QCD photons at RHIC energy. $p_T$
broadening effect is ignored. At RHIC energy the effect is not as important
as at SPS energy. The dotted lines are the thermal photons obtained for
initial radial velocity $v_r^0$=0,0.3 and 0.5c. Solid lines show the total
contribution ( thermal and hard QCD photons). It can be seen that for $p_T <$
2 GeV, thermal photons dominate the spectrum. Hard QCD photons dominate at
higher $p_T >$ 4 GeV. In the intermediate $p_T$ range (2 GeV-4 GeV), thermal
photons contribute as copiously as the hard QCD photons. This is the range
where initial fluid velocity has considerable effect. Considering the
importance of hard QCD photons at RHIC energy, it is of utmost importance to
determine experimentally the single photon spectra in pp collisions at RHIC
energy. Only then it will be possible to comment on the possible QGP
formation at RHIC energy from the single photon data.

To summarize, we have analyzed the recent WA98 single photon data assuming
QGP formation in the initial state. It was shown that in order to have a
reliable estimate of photon yield, initial time of the hydrodynamic
evolution should be larger than 0.4 fm, otherwise neglect of three and
higher loops in the photon production rate

\begin{figure}[h]
\centerline{\psfig{figure=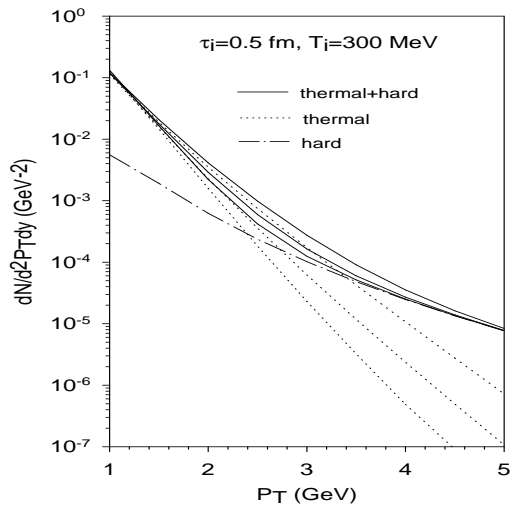,height=7.5cm,width=7.5cm}}
\caption{Theoretical estimate of direct photons at RHIC energy for Au+Au
collisions at ${\ss }$=130 GeV. The dash-dot line is the hard QCD photons.
The dotted lines are the thermal photons with initial radial velocity $v_r^0$%
=0,0.3 and 0.5c. The solid lines are the total contribution of hard QCD
photons and thermal photons.}
\end{figure}

\noindent would be unjustified. They can contribute in the same order as the
two loop photon rate (which is more than 70\% at high $p_{T}$ for an initial
time of 0.2 fm). As far as $\tau _{i}$ $\geq $ 0.4 fm, then initially static
QGP scenario underestimates the high $p_{T}$ photons compared to the
observed data. The data could be fitted only if some kind of initial radial
velocity is present. Photon yield depend on the initial velocity
distribution, changing by factor of 2-3 at the high $p_{T}$ , from the
surface emission scenario (surface peaked) to the detonation at the center
(Woods-Saxon form). WA98 data could be fitted reasonably well if the
thermalisation time $\tau _{i}$, ranges between 0.4-1.0 fm (corresponding to
initial temperature between 271-200 MeV), and if the QGP fluid has an
initial (surface peaked) radial velocity in the ranges of 0.3-0.5c. The
model was applied to predict photon spectra at RHIC energy in Au+Au
collisions. It was shown that high $p_{T}$ spectra is dominated by hard QCD
photons. Thermal photons dominate low $p_{T}$ part of the spectra and in the
intermediate range of $p_{T}$ thermal photons contribute comparably to hard
photons. We feel it is important to measure hard QCD photons in pp
collisions at RHIC energy. 
Only then it will then be possible to comment on possible QGP formation
from the direct photon data at RHIC.

This work is supported in part by CNPq/MCT and FAPERJ. One of the author
(AC) thanks the Lady Davis fellowship trust for supporting the visit to
Technion, where part of the work is done.

\begin{table}[tbp]
\caption{The initial temperature of the QGP for certain initial times $%
\protect\tau_i$. dN/dY was assumed to be 750.}
\label{table1}%
\begin{tabular}{cccccc}
$\tau_i(fm)$ & 0.2 & 0.4 & 0.6 & 0.8 & 1.0 \\ 
$T^{QGP}_i (MeV)$ & 341 & 271 & 237 & 215 & 200%
\end{tabular}%
\end{table}


\begin{references}
\bibitem[*]{byline} e-mail address:akc@veccal.ernet.in

\bibitem{QM2001} Proceedings of Quark Matter 2001,

\bibitem{he00} U. Heinz and M. Jacob, nucl-th/0002042 .

\bibitem{Zsch} D. Zschiesche et al, nucl-th/0101047, Proceedings of the
Symposium on Fudamental Issues in Elementary Matter, ed. W.Greiner,Debrecen,
Hungary

\bibitem{wa98} M. M. Aggarwal et al, WA98 collaboration, nucl-ex/0006008,
Phys. Rev. Lett. {\bf 85},3595 (2000).

\bibitem{sr00} D. K. Srivastava and B. Sinha, nucl-th/0006018, Phys. Rev.
C64, 034902 (2001).

\bibitem{ja00} Jan-e Alam, S. Sarkar, T. Hatsuda, T. K. Nayak and B. Sinha,
Phys. Rev. C63, 021901 (2001).

\bibitem{ga00} K. Gallmeister, B. Kampfer and O. P. Pavlenko, Phys. Rev.
C62, 057901 (2000).

\bibitem{pe00} D.Y. Peressounko and Yu. E. Pokrovsky, hep-ph/0002068.

\bibitem{ch01} A. K. Chaudhuri, nucl-th/0012058.

\bibitem{ge86} H. von Gersdorfff, M. Kataja, L. McLerran and P. V.
Ruuskanen, Phys. Rev. D{\bf 34},794 (1986).

\bibitem{hw85} R. C. Hwa and K. Kajantie, Phys. Rev. D{\bf 32}, 1109 (1985) .

\bibitem{na92} H. Nadeau, J. Kapusta and P. Lichard, Phys. Rev. C {\bf 45}%
3034 (1992).

\bibitem{xi92} L. Xiong, E. Shuryak and G. E. Brown, Phys. Rev. D{\bf 46}%
3798 (1992) .

\bibitem{ka91} J. Kapusta, P. Lichard and D. Seibert, Phys. Rev. D{\bf 44}
2774 (1991).

\bibitem{au98} P. Aurenche, F. Gelis, H. Zaraket and R. Kobes, Phys. Rev.D 
{\bf 58},085003 (1998).

\bibitem{st01} F. D. Steffen and M. A. Thoma, hep-ph/0103044, Phys.Lett.B%
{\bf 510}98 (2001).

\bibitem{au98a} P. Aurenche, private communication.

\bibitem{au00a} P. Aurenche, F. Gelis, H. Zaraket, Phys. Rev.D{\bf 61}%
,116001,2000.

\bibitem{au00b} P. Aurenche, F. Gelis, H. Zaraket, Phys. Rev.D{\bf 62}%
,096012,2000.

\bibitem{du01} A. Dumitru, L. Frankfurt, L. Gerland, H. Stocker and M.
Strikman, hep-ph/0103203, Phys.Rev.C{\bf 64} 054909(2001).

\bibitem{tr96} C. T. Traxler and M. H. Thoma, Phys. Rev. C53(1996)1348.

\bibitem{ro97} P. Roy, J. Alam, S. Sarkar, B. Sinha and S. Raha, Nucl. Phys.
A624 (1997)687.

\bibitem{du99} A. Dumitru, D. H. Rischke, Phys. Rev. C59, 354 (1999).
\end{references}
\end{document}